\documentstyle[multicol,overcite,pre,aps,epsf]{revtex}

\begin{document}
\tightenlines

\title{Emergence of Quantum-Classical Dynamics in an Open Quantum Environment}
\author{Kazutomu Shiokawa and Raymond Kapral
\\ Chemical Physics Theory Group, Department of Chemistry,\\
University of Toronto, Toronto, ON M5S 3H6, Canada}
\maketitle

\begin{abstract}
The conditions under which an open quantum mechanical system may
be described by mixed quantum-classical dynamics are investigated.
Decoherence is studied using influence functional methods in a model
composite quantum system comprising two coupled systems, $A$ and $C$,
interacting with a harmonic bath with Ohmic and
super-Ohmic spectral densities. Subsystem $A$ is directly coupled
to subsystem $C$, while $C$ is coupled directly to the bath.
Calculations are presented for a model where subsystem $A$ is
taken to be a two-level system which is bilinearly coupled to a
single harmonic oscillator $C$ subsystem. The loss of quantum
coherence in each subsystem is discussed in the extreme
non-adiabatic regime where the dynamics of subsystem $A$ is
essentially frozen. 
Subsystem $C$ is shown to lose its coherence rapidly, while
subsystem $A$ maintains coherence for longer time periods since
$C$ modulates the influence of the bath on $A$. Thus,
one may identify situations where the coupled $AC$ system
evolution effectively obeys mixed quantum-classical dynamics.
\end{abstract}
\pacs{xx}
\begin{multicols}{2}
\narrowtext

\section{Introduction}

Since it is often impossible to isolate a quantum system
completely from its environment, the study of open quantum
dynamical systems is necessary in many circumstances. Techniques
used to treat open quantum systems have been developed extensively
and used in numerous applications. \cite{davis,Mukamel95,Weiss99}
Projection operator techniques \cite{Nakajima58,Zwanzig61} and
influence functional methods \cite{FeyVer63} have been
actively used in chemical physics in studies of electron, proton,
and exciton transfer processes in the condensed phase and in
biological systems. \cite{QDOS,makri,EgoEveSki99,MayKuhn00}

In this paper, we investigate some of the circumstances under
which an open quantum system can be described using mixed
quantum-classical dynamics. Quantum-classical dynamics is
used to study condensed phase many-body systems
\cite{tully0,herman}, especially in the context of methods for
treating non-adiabatic dynamics
\cite{tully1,coker,rossky,martinez,prezhdo,martens,kapral,schofield}.
In the studies presented here, we consider a composite quantum
mechanical system $AC$ comprising two coupled subsystems $A$ and
$C$; subsystem $C$ is assumed to be in direct contact with a
thermal quantum mechanical bath $B$. Systems with this structure
can arise in condensed phase dynamics where certain quantum
degrees of freedom ($A$), for example, those associated with
protons or electrons, may interact directly with neighboring
solvent molecules ($C$), which in turn interact with the rest of
the solvent ($B$). It is  interesting to determine if
the composite system may be treated as
mixed quantum-classical system where the dynamics of subsystem
$C$, which is in direct contact with the heat bath, is classical
in character while subsystem $A$ retains its quantum nature. Some
aspects of the dynamics of such quantum-classical composite systems
have been investigated.  \cite{kapral2}

In order to investigate this problem we study a simple model
system where subsystem $A$ depends on spin degrees of freedom and
subsystem $C$ is a single harmonic oscillator bilinearly coupled
both to subsystem $A$ and the bath. The bath is a collection of
independent harmonic oscillators. While this is a highly
simplified model of the realistic systems discussed above, it does
capture some essential features of real coupled systems and is
amenable to detailed analysis. Due to its interaction with the
bath, the dynamics of subsystem $C$ is dissipative and executes
brownian motion. The brownian motion of a quantum particle
governed by different potential functions and immersed in a
thermal harmonic oscillator bath has been studied
extensively by influence functional methods\cite{CaldeiraLeggett84,
HakimAmb85,GSI88,UnruhZurek89,QBM1,QBM2}.
The character of its dynamics is
determined by the system-bath coupling and the spectral density
and temperature of the bath. In the composite system we study, the
dynamics of the $C$ subsystem oscillator is also influenced by the
quantum dynamics of subsystem $A$. The dynamics of subsystem $A$
is more complicated. It is also dissipative but its energy must be
transmitted through subsystem $C$ to the bath. Our results on the
applicability of mixed quantum-classical dynamics are based on the
nature of decoherence\cite{EID,Dec96} in the coupled system: when one subsystem
behaves quantum mechanically and the other classically, there must
be a mechanism making the former decohere slowly and the latter
quickly.

The outline of this paper is as follows: In Sec.~\ref{sec:IF} we
specify the model in detail and outline the application of the
influence functional formalism to it. Although we
use influence functional methods, similar results can be obtained
by other methods. Section~\ref{sec:QBM} considers the
equilibration of the $CB$ system in the absence of $A$ and
establishes that this composite system can be described by an
effective spectral density. The emergence of mixed
quantum-classical dynamics is investigated in
Sec.~\ref{sec:emergence}. In this section we use the results
developed in the previous sections, specialized to a two-level system
in the strongly non-adiabatic regime where the populations change
very slowly. We study the decoherence of the off-diagonal elements
of the subsystem $A$ reduced density matrix in this regime and
show under what conditions one may observe quantum-classical
dynamics. The conclusions of the study are presented in
Sec.~\ref{sec:conc}. The Appendices contain additional details of
the calculations for Ohmic and super-Ohmic spectral densities.

\section{General formulation} \label{sec:IF}

\subsection{Composite system in a thermal bath}

The composite $ACB$ quantum system consists of three coupled subsystems:
subsystem $A$ is coupled directly to subsystem $C$, while subsystem $C$
is in direct contact with a thermal bath $B$. The $ACB$ system hamiltonian
is,
\begin{eqnarray}
\hat{H} = \hat{H}_A + \hat{V}_{AC} + \hat{H}_C + \hat{V}_{CB}
+\Delta \hat{H}_C + \hat{H}_B\;.
\label{H}
\end{eqnarray}
The thermal bath is taken to be a set of $N$ independent harmonic
oscillators with frequencies $\omega_{n}$, masses $m_{n}$, and coordinates
and conjugate momenta
$(\hat{q},\hat{p})\equiv (\hat{q}_1,...,\hat{q}_N,\hat{p}_1,...,\hat{p}_N)$
so that the hamiltonian is
\begin{eqnarray}
 \hat{H}_{B} = \sum_{n=1}^{N}( \frac{\hat{p}_{n}^{2}}{2 m_{n}}
    + \frac{1}{2}m_{n} \omega_{n}^{2}\hat{q}_{n}^{2}).
     \label{HB}
\end{eqnarray}
Subsystem $C$ consists of a single harmonic oscillator with renormalized
mass $M$ and frequency $\omega_r$ with coordinate  and momentum operators
$(\hat{x}, \hat{p})$,
\begin{eqnarray}
\hat{H}_C = \frac{\hat{p}^2}{2M} + \frac{M}{2} \omega_r^2 \hat{x}^2\;.
\label{HC}
\end{eqnarray}
The interaction potential between $C$ and the bath is assumed to be bilinear,
\begin{eqnarray}
 \hat{V}_{CB} = \hat{x} \sum_{n=1}^{N} c_{n} \hat{q}_{n},
    \label{HI}
\end{eqnarray}
where $c_n$ is the coupling constant to the $n$th oscillator. The coupling
constants are related to the spectral density of the bath by,
\begin{eqnarray}
J_B(\omega)
\equiv \pi \sum_{n}\frac{c_{n}^2}{2 m_n \omega_n} \delta(\omega - \omega_n).
\label{SpectralDensity}
\end{eqnarray}
We suppose the spectral density has the form
\begin{equation}
J_{B}(\omega) = \eta \omega^{\nu} e^{-\omega/\Lambda},
\end{equation}
and consider both Ohmic ($\nu=1$) and super-Ohmic ($\nu=3$) cases.
A counter term $\Delta \hat{H}_C$
has been added to the hamiltonian. It depends on $c_n$, $m_n$,
$\omega_n$, $\hat{p}$ and $\hat{x}$ and is given by
\begin{eqnarray}
 \Delta \hat{H}_c &=&
\left\{
\begin{array}{lll}
\eta \Lambda \hat{x}^2 / \pi  & \hspace{1cm} &(\nu=1) \\
\eta \Lambda \hat{p}^2 / M^2 \pi  + \eta \Lambda^3 \hat{x}^2 / \pi
&\hspace{1cm} &(\nu=3).
\end{array}
\right.
\label{counterterm}
\end{eqnarray}
This term is introduced to
cancel the shift of the mass and frequency of the $C$ subsystem
oscillator due to the interaction with the bath which will
become divergent when the frequency cutoff $\Lambda$ goes to infinity.
As is customary, we consider the
renormalized quantities after including a counter term
as physical observables with specified values in our analysis \cite{IZ}.
The hamiltonian of subsystem $A$, $\hat{H}_A(\sigma)$, depends on
spin variables $\sigma$ and we assume subsystem $A$ is bilinearly coupled
to subsystem $C$, $\hat{V}_{AC}=\lambda \sigma \hat{x}$.

The dynamical properties of interest can be computed from the
density matrix of the system at time $t$. In the coordinate
representation $\{x,q\}$ of the $C$ subsystem and bath $B$, the
density matrix takes the form, $\hat{\rho}(x,x',q,q',t) \equiv
\langle x ~q | \hat{\rho}(t) | x' ~q' \rangle$, which is still an
operator in the $A$ subsystem degrees of freedom. The density
matrix may be written more explicitly in integral form in terms of
the propagator
\begin{eqnarray}
       \hat{K}(x,q;t \mid x_0, q_{0};0)
 \equiv \langle x ~q | e^{- i \hat{H} t/ \hbar} | x_0 ~q_{0}  \rangle \;,
          \label{KernelK}
\end{eqnarray}
as
\begin{eqnarray}
  \hat{\rho}(x,x',q,q',t) &=&
       \int d x_{0} d x'_{0} d q_{0} d q'_{0}
       \hat{K}(x,q; t \mid x_0, q_{0};0)  \nonumber \\
&\times&  \hat{\rho}(x_0, x'_0, q_{0}, q'_{0}, 0)
   \hat{K}^{*}(x',q'; t \mid x'_0, q'_{0};0) \;. \nonumber \\
          \label{DMintegral}
\end{eqnarray}

We are primarily interested in the dynamics of the composite $AC$
subsystem under the influence of the thermal bath.  In such a
circumstance the reduced density matrix $\hat{\rho}_r$, obtained by
integrating over the bath degrees of freedom, is the relevant
quantity. Such a reduction is especially appropriate if the
characteristic time scale for the bath is much shorter than those
for the $A$ and $C$ subsystems. We assume a factorized initial
condition between the $AC$ subsystem and the bath,
\begin{eqnarray}
  \hat{\rho}(x_0, x'_0, q_{0}, q'_{0}, 0) =
  \hat{\rho}_{AC}(x_0, x'_0, 0)
  \otimes
  \rho_{B}(q_{0}, q'_{0},0)\;,
          \label{IDMAC}
\end{eqnarray}
and the bath is always taken to be initially in thermal equilibrium.
Under these conditions the integral form of the reduced density matrix
at time $t$ is
\begin{eqnarray}
  \hat{\rho}_r(x,x',t) &=&
       \int d x_{0} d x'_{0}
       \hat{J}_r(x, x';t \mid x_0, x'_0; 0)
       \hat{\rho}_{AC}(x_0, x'_0, 0), \nonumber \\
            \label{DMredintegral}
\end{eqnarray}
where the time evolution kernel $J_r$ is given by
\begin{eqnarray}
 \hat{J}_r(x, x';t \mid x_0, x'_0; 0) &=&
  \int d q d q_{0} d q'_{0}
  \hat{K}(x,q;t \mid x_0, q_{0};0) \nonumber \\
&\times&   \rho_{B}(q_{0}, q'_{0} ,0)
  \hat{K}^{*}(x',q; t \mid x'_0, q'_{0};0). \nonumber \\
   \label{Jr}
\end{eqnarray}
Given the initial conditions discussed above, this exact
expression for the reduced density matrix specifies a
non-Markovian time evolution since the solution at time $t$
depends on its past history. Approximate Markovian evolution equations
may miss essential features of the quantum/classical correspondence and
tend to underestimate the loss of quantum coherence. In addition, a Markovian
approximation is not generally valid for a harmonic oscillator
model, except for systems with Ohmic type dissipation in the high
temperature regime.
Below we use
influence functional methods where the exact solution is expressed
as a path integral in order to deal with the nonlocal time
evolution.

\subsection{Influence functional method}

The formulation of the reduced density matrix in terms of an
influence functional for the composite $ACB$ system parallels
that for spin-boson models often discussed in the literature.
\cite{LCDFGZ87,Weiss99} Consequently the presentation of the
generalization to our system will be brief.
In order to write the reduced density matrix in terms of an
influence functional, we first introduce a basis of spin
functions with labels $\sigma$ and $\sigma'$ to represent the states of
the $A$ subsystem. In this basis we may write eq.~(\ref{DMredintegral}) as,
\begin{eqnarray}
    && \rho_r(\sigma, \sigma', x, x', t)=  \int d \sigma_0 d \sigma'_0
            \int d x_{0} d x'_{0} \nonumber \\
       && \times J_r(\sigma, \sigma', x, x';t
             \mid \sigma_0, \sigma_0', x_0, x'_0; 0)
             \rho_{r}( \sigma_0, \sigma_0', x_0, x'_0, 0). \nonumber \\
\label{DMACf}
\end{eqnarray}
For the factorized initial condition (\ref{IDMAC}) the kernel
$J_r$ can be written in the path integral form as
\begin{eqnarray}
&& J_r(\sigma, \sigma', x, x';t \mid \sigma_0, \sigma'_0, x_0, x'_0; 0) = \nonumber \\
&&  \int^{(\sigma \sigma' )}_{(\sigma_0 \sigma'_0 )} {\mathcal D}
\sigma {\mathcal D} \sigma'  K[\sigma] K^{*}[\sigma']
  J_{C}^{\sigma \sigma'}(x, x';t \mid x_0, x'_0; 0),
  \label{Jrpathint}
\end{eqnarray}
where $K[\sigma]$ is a probability amplitude for subsystem $A$
following the path $\sigma$ in the absence of the subsystem $C$
and the environment. The evolution kernel $J_{C}^{\sigma \sigma'}$ for
subsystem $C$ in the presence of external sources $(\sigma,
\sigma')$ is
\begin{eqnarray}
&&J_{C}^{\sigma \sigma'}(x, x';t \mid x_0, x'_0; 0)
  \equiv \int^{(x x')}_{(x_0 x'_0)} {\mathcal D} x {\mathcal D} x'
  e^{i {\cal S}[\sigma,\sigma',x,x']/\hbar},
\nonumber \\
\label{JF}
\end{eqnarray}
The action ${\cal S}[\sigma,\sigma',x,x']$ consists of several
contributions and can be decomposed as follows:
\begin{eqnarray}
{\cal S}[\sigma,\sigma',x,x']&=& {\cal S}_C[x,x'] +
{\cal S}_{AC}[\sigma,\sigma',x,x']   \nonumber \\
&&+\Delta {\cal S}_C[x,x']
+{\cal S}_{IF}[x,x'].
\end{eqnarray}
The actions for subsystem $C$, plus its counter action, and the
interaction between subsystems $A$ and $C$ are
\begin{eqnarray}
({\cal S}_C+\Delta {\cal S}_C)[x,x']&=& \int_{0}^{t} ds \Big(
\frac{M_{0}}{2}  \dot{x}^2(s) -
\frac{M_{0}}{2}\omega_{0}^2 x^2(s) \nonumber \\
&&-\frac{M_{0}}{2} \dot{x'}^2(s) +
\frac{M_{0}}{2}\omega_{0}^2 x^{\prime 2}(s) \Big), \nonumber \\
{\cal S}_{AC}[\sigma,\sigma',x,x']&= &
\lambda \int_{0}^{t} ds \left( x(s) \sigma(s) - x'(s) \sigma'(s)
\right)\;, \nonumber \\
\label{SAC}
\end{eqnarray}
where, for notational convenience, we used the bare mass $M_{0}$ and bare frequency $\omega_{0}$ as
$M_{0}=M$ and $M_{0}\omega_{0}^2=M\omega_r^2+
2\eta \Lambda/\pi$ for $\nu=1$ while $M_{0}=M+2\eta \Lambda/\pi$
and $M_{0}\omega_{0}^2=M\omega_r^2+ 2\eta \Lambda^3/(3\pi)$
for $\nu=3$.

The influence action ${\cal S}_{IF}[x,x']$ accounts for the effect
of the bath on $C$ and is given by
\begin{eqnarray}
{\cal S}_{IF}[x,x'] &=&
i \int_{0}^{t} ds \int_{0}^{s} ds'
[ x(s) - x'(s) ] \times \nonumber \\
&&[\alpha(s-s') x(s') - \alpha^{*}(s-s') x'(s')],
\label{IFaction}
\end{eqnarray}
where $\alpha(t)$ is a complex kernel whose real $\alpha^R(t)$
and imaginary $\alpha^I(t)$ parts, respectively, are given by
\begin{eqnarray}
\alpha^R(t)&=& \frac{1}{\pi} \int_{0}^{\infty} d \omega J_B(\omega)
\coth \frac{\beta \hbar \omega}{2} \cos \omega t, \\
\alpha^I(t)&=&-\frac{1}{\pi} \int_{0}^{\infty} d \omega J_B(\omega)
\sin \omega t.
\label{noisekernel}
\end{eqnarray}
We define new variables as $R \equiv (x+x')/2$,
$r \equiv x-x'$, $\sigma_{\pm} \equiv \sigma \pm \sigma'$, and
write the actions in these new variables as
\begin{eqnarray}
&&{\cal S}_{AC}[\sigma_{\pm},R,r]= \frac{\lambda}{2}
\int_{0}^{t} ds \left( \sigma_{+}(s) r(s) + 2 \sigma_{-}(s) R(s)
\right), \nonumber \\
&&({\cal S}_{C}+\Delta {\cal S}_C)[R,r]= \int_{0}^{t} ds \{ M_{0}
\dot{R}(s)\dot{r}(s) - M_{0}\omega_{0}^2 R(s)r(s) \}, \nonumber \\
&&{\cal S}_{IF}[R,r] = i \int_{0}^{t} ds \int_{0}^{s} ds' r(s)
\alpha^R(s-s') r(s') \nonumber \\
&&\qquad \qquad - 2 \int_{0}^{t} ds \int_{0}^{s} ds' r(s)
\alpha^I(s-s') R(s').
\label{SAC2}
\end{eqnarray}

\subsection{Euler-Lagrange equations and solution} \label{EL}

>From eqs.~(\ref{SAC2}) the Euler-Lagrange equations for $R$ and $r$ are
\begin{eqnarray}
 M_{0} \ddot{R}_c(s)
+  M_{0} \omega_{0}^2 R_c(s)
&+& 2 \int_{0}^{s} ds'
\alpha^I(s-s') R_c(s') \nonumber \\
&=& \frac{1}{2}\lambda \sigma_{+}(s),
\label{EL1}
\end{eqnarray}
\begin{eqnarray}
M_{0} \ddot{r}_c(s) +
 M_{0} \omega_{0}^2 r_c(s)
&-& 2 \int_{s}^{t} ds' \alpha^I(s-s') r_c(s') \nonumber \\
&=&\lambda \sigma_{-}(s). \label{EL2}
\end{eqnarray}
The initial and final conditions for eq.~(\ref{EL1})
(eq.~(\ref{EL2})) are $R_0$ and $R_t$ ($r_0$ and $r_t$).
If we let the two independent solutions of the homogeneous part of
eq.~(\ref{EL1}) (eq.~(\ref{EL2})) be $u_i(s)$($v_i(s)$), $i=1,2$,
with boundary conditions
$u_1(0)=1,u_1(t)=0$, $u_2(0)=0,u_2(t)=1$
($v_1(0)=1,v_1(t)=0$, $v_2(0)=0,v_2(t)=1$),
the solutions of these uncoupled equations can be written as
\begin{eqnarray}
R_c(s)
&=& R_0 u_1(s)  + R_t u_2(s)
+\frac{\mu}{2} \sigma_{+}(g(s)),
\nonumber \\
r_c(s)
&=& r_0 v_1(s)  + r_t v_2(s)
+ \mu \sigma_{-}(g(s)),
\label{SolutionRr}
\end{eqnarray}
where we have used the notation,
\begin{eqnarray}
 \sigma_{+}(g(s)) &\equiv& \int_{0}^{s} ds' g_{+}(s-s')
 \sigma_{+}(s') \nonumber \\
&-& u_2(s) \int_{0}^{t} ds' g_{+}(t-s') \sigma_{+}(s'),
 \nonumber \\
 \sigma_{-}(g(s)) &\equiv& \int_{0}^{s} ds' g_{-}(s-s')
 \sigma_{-}(s') \nonumber \\
 &-& v_2(s) \int_{0}^{t} ds' g_{-}(t-s') \sigma_{-}(s').
\label{g+g-}
\end{eqnarray}
and defined $\mu \equiv \lambda / M$.
The solutions $v_1$ and $v_2$ satisfy the homogeneous part of
the backward time equation (\ref{EL2})
and are related to $u_1$ and $u_2$ by $v_1(s)=u_2(t-s)$ and
$v_2(s)=u_1(t-s)$. The function
$g_{+}(s)$ ($g_{-}(s)$) also satisfies the homogeneous part of
eq.~(\ref{EL1}) (eq.~(\ref{EL2}))
with boundary conditions $g_{\pm}(0)=0,\dot{g}_{\pm}(0)=1$.

The solutions for $g_{\pm}$ are given in Appendix A for Ohmic
and super-Ohmic spectral densities. From these solutions $u_{1,2}$
and $v_{1,2}$ can be determined.

\subsection{Reduced density matrix solution}

Since the potentials in our model are harmonic, an exact evaluation of the path integral
can be carried out. It is dominated by the classical solution given
in eq.~(\ref{SolutionRr}). The $\sigma$ dependence in eq.~(\ref{SolutionRr})
arises from the back reaction of the $A$ subsystem on the $C$
subsystem. To separate out the contribution from this back reaction,
we expand ${\cal S}$ in powers of $\sigma$ and write,
\begin{eqnarray}
{\cal S}[\sigma_{\pm},R_c,r_c]&=&
{\cal S}^{(0)}[R_c,r_c] +
{\cal S}^{(1)}[\sigma_{\pm},R_c,r_c] +
{\cal S}^{(2)}[\sigma_{\pm}]. \nonumber \\
\label{S-SSS}
\end{eqnarray}
The zeroth order term in $\sigma$ takes the form
\begin{eqnarray}
{\cal S}^{(0)}[R_c,r_c]
&=& \Big( M \dot{u}_1(t) R_0 + M \dot{u}_2(t) R_t\Big) r_t \nonumber \\
&-& \Big( M \dot{u}_1(0) R_0 +  M \dot{u}_2(0) R_t\Big)r_0 \nonumber \\
&+& i \Big( a_{11}(t) r_0^2 + ( a_{12}(t) + a_{21}(t)) r_0 r_t
\nonumber \\
&&+ a_{22}(t) r_t^2\Big)\;,
\label{SQBM}
\end{eqnarray}
and is an influence action for a damped harmonic
oscillator\cite{GSI88,QBM1}. Here
\begin{eqnarray}
a_{kl}(t) &=& \frac{1}{2} \int_{0}^{t} ds \int_{0}^{t} ds'
v_k(s) \alpha^R(s-s') v_l(s'),
\label{aij}
\end{eqnarray}
for $(k,l=1,2)$ contains the effects of fluctuations due to the bath on
subsystem $C$. The term linear in $\sigma$ is
\begin{eqnarray}
&&{\cal S}^{(1)}[\sigma_{\pm},R_c,r_c] = \lambda \int_{0}^{t} ds
\left( u_1(s) R_0 + u_2(s) R_t \right) \sigma_{-}(s) \nonumber \\
&+& i \mu r_0 \int_{0}^{t} ds \int_{0}^{t} ds'
\{ v_1(s) \alpha^R(s-s') \sigma_{-}(g(s')) \} \nonumber \\
&+& i \mu r_t \int_{0}^{t} ds \int_{0}^{t} ds'
\{ v_2(s) \alpha^R(s-s') \sigma_{-}(g(s')) \} \nonumber \\
&+& \frac{\lambda r_0}{2}  \int_{0}^{t} ds \;
v_1(s) \sigma_{+}(s)  +  \frac{\lambda r_t}{2}\int_{0}^{t} ds
v_2(s) \sigma_{+}(s)\;.
\label{SABC}
\end{eqnarray}
This term arises from the interaction between subsystem $A$ and
the bath, modulated by the trajectory of subsystem $C$.
The quantum back reaction of subsystem $A$ on $C$ induces
self-coupling in $A$, which is contained in the last term,
\begin{eqnarray}
{\cal S}^{(2)}[\sigma_{\pm}]
&=& \frac{\lambda \mu}{2}
\int_{0}^{t} ds \; \sigma_{-}(s) \sigma_{+}(g(s)) \nonumber \\
&+&i \mu^2 \int_{0}^{t} ds \int_{0}^{s} ds' \;
\sigma_{-}(g(s)) \alpha^R(s-s') \sigma_{-}(g(s')). \nonumber \\
\label{SAB}
\end{eqnarray}

Using the results above, $J_C^{\sigma \sigma'}$ in eq.~(\ref{JF})
can be written in the compact form,
\begin{eqnarray}
J_{C}^{\sigma \sigma'}(R_t, r_t;t  \mid R_0, r_0; 0)   =
     N(t) \exp\Big\{\frac{i}{\hbar} {\mathcal L} \Big\},
  \label{Jrsimple}
\end{eqnarray}
where ${\mathcal L}= \vec{R}^{T} {\bf u} \vec{r} +i \vec{r}^{T}{\bf a} \vec{r}
+ \vec{\sigma}_{u}^{T} \vec{R} + \vec{\sigma}_{v}^{T} \vec{r}
+ {\cal S}^{(2)}[\sigma_{\pm}] $ and $N(t)$ is a normalization factor
independent of dynamical variables. In writing this equation we have
introduced the notation and $({\bf a})_{ij} = a_{ij}$,
$\vec{R}^T=(R_0, R_t)$ and $\vec{r}^T=(r_0, r_t)$, where $T$
stands for the transpose. The matrix ${\bf u}$ is defined as
\begin{eqnarray}
{\bf u} =
        \left( \begin{array}{cc}
       u_{11}  & u_{12}   \\
       u_{21}  & u_{22}
         \end{array}      \right)
 \equiv
       M  \left( \begin{array}{cc}
       - \dot{u}_{1}(0) & \dot{u}_{1}(t)  \\
       - \dot{u}_{2}(0) & \dot{u}_{2}(t)
         \end{array}      \right)\;,
\label{U}
\end{eqnarray}
while the vectors $\vec{\sigma}_{u}$ and $\vec{\sigma}_{v}$ are given by
\begin{eqnarray}
\vec{\sigma}_{u} =
\lambda \left( \begin{array}{c}
        \sigma_{-}(u_1) \\
        \sigma_{-}(u_2)
       \end{array}      \right), \quad
\vec{\sigma}_{v} =
\lambda \left( \begin{array}{c}
        \sigma(v_1) \\
        \sigma(v_2)
       \end{array}      \right)\;.
\label{Sigmauv}
\end{eqnarray}
In eqs.~(\ref{Sigmauv}) we have used the symbolic notation
\begin{eqnarray}
\sigma_{-}(u_i) &\equiv& \int_{0}^{t} ds u_i(s) \sigma_{-}(s) \;, \nonumber \\
\sigma(v_i) &\equiv& \int_{0}^{t} ds v_i(s) [ \sigma_{+}(s)/2 \nonumber \\
&&+ \int_{0}^{t} ds'   i \alpha^R(s-s') \sigma_{-}(g(s'))/M]\;,
\end{eqnarray}
for $i=1,2$.

Having derived the expression for $J_C^{\sigma \sigma'}$ in
eq.~(\ref{Jrsimple}), we may use it to obtain the partial Wigner
transform of the reduced density matrix. The density matrix in the
partial Wigner representation \cite{Wigner} over the $C$ subsystem
degrees of freedom is defined by
\begin{eqnarray}
\rho_{rW}(\sigma,\sigma',R,P,t)&=& \frac{1}{2\pi \hbar} \int dr
e^{-iPr/\hbar} \nonumber \\
&& \times \rho_r(\sigma,\sigma',R+r/2,R-r/2,t).
\end{eqnarray}
Wigner transforming eq.~(\ref{DMACf}) and using this definition we
obtain
\begin{eqnarray}
\rho_{rW}(\sigma,\sigma',R,P,t)&=&\int d\sigma_0 d\sigma_0' \;
\int^{(\sigma \sigma' )}_{(\sigma_0 \sigma'_0 )} {\mathcal D}
\sigma {\mathcal D} \sigma'  K[\sigma] K^{*}[\sigma']\nonumber \\
&& \int dR_0 dP_0 \; K_{C}^{\sigma \sigma'}(R, P;t \mid R_0, P_0;
0)\nonumber \\
&&\times \rho_{rW(}\sigma_0,\sigma'_0,R_0,P_0,0) \;,
\label{Wrdensity}
\end{eqnarray}
where the kernel $K_{C}^{\sigma \sigma'}(R, P;t \mid R_0,P_0;0)$
is defined by
\begin{eqnarray}
K_{C}^{\sigma \sigma'}(R, P;t \mid R_0,P_0;0)&=& \frac{1}{2 \pi
\hbar} \int dr dr_0\; e^{-i(Pr-P_0r_0)/\hbar} \nonumber \\
&& \times J_{C}^{\sigma \sigma'}(R, r;t  \mid R_0, r_0; 0)\;.
\end{eqnarray}

Inserting the expression for $J_{C}^{\sigma \sigma'}$ in eq.~(\ref{Jrsimple}), the
time evolution kernel $K_{C}^{\sigma \sigma'}$ for the Wigner function is given by
\begin{eqnarray}
 K_{C}^{\sigma \sigma'}(R_t, P_t; t &\mid& R_0, P_0; 0)  \nonumber \\
&=& \frac{N(t)}{2 \pi \hbar} \int dr dr_0 \;
e^{i (-P_t r_t + P_0 r_0+ {\mathcal L})/\hbar} \nonumber \\
 &=& N_W(t) \exp\Big[ - \delta \vec{X}_{W}^{T}
{\bf Q}^{-1} \delta \vec{X}_{W} \nonumber \\
&&+ \frac{i}{\hbar} \vec{\sigma}_{u}^{T} \vec{R}
+ \frac{i}{\hbar}{\cal S}^{(2)}[\sigma_{\pm}] \Big],
  \label{Ksigma}
\end{eqnarray}
where $N_W(t)=N(t)/2 \hbar \sqrt{|{\bf a}|}$ and $|{\bf a}|$ is the
determinant of ${\bf a}$. The vector $\delta \vec{X}_W=
\vec{X}_W -\langle \vec{X} \rangle$, with
\begin{eqnarray}
\vec{X}_{W} =
\left( \begin{array}{c}
        R_t \\
        P_t - \lambda \sigma(v_2)
       \end{array}      \right) \;,
\label{RWvec}
\end{eqnarray}
and
\begin{eqnarray}
\langle \vec{X} \rangle=
\left( \begin{array}{c}
        \langle R \rangle \\
        \langle P \rangle
         \end{array}      \right)
=\frac{-1}{u_{21}}
   \left( \begin{array}{cc}
       u_{11} & 1 \\
       |{\bf u}|  & u_{22}
         \end{array}      \right)
\left( \begin{array}{c}
        R_0 \\
        P_0 + \lambda \sigma(v_1)
         \end{array}      \right).
\label{RC}
\end{eqnarray}
The matrix ${\bf Q}$ is given by
\begin{eqnarray}
&&{\bf Q} = \frac{4\hbar}{u_{21}^2} \times \nonumber \\
&&
       \left( \begin{array}{cc}
        a_{11}                       & a_{12} u_{21} - a_{11} u_{22} \\
       a_{12} u_{21} - a_{11} u_{22} &
       a_{11} u_{22}^2 - 2 a_{12} u_{21} u_{22} + a_{22} u_{21}^2
         \end{array}      \right). \nonumber \\
\label{QW2}
\end{eqnarray}
In the next section we shall need $Q_{11}$ and the
explicit expression for this matrix element is
\begin{eqnarray}
Q_{11}(t)&=&  \frac{2 \hbar}{\pi M^2}  \int_{0}^{\infty} d \omega
 \coth \left( \frac{\beta \hbar \omega}{2} \right)
J_B(\omega) \Delta(\omega,t), \nonumber \\
\label{Q11}
\end{eqnarray}
where $\Delta(\omega,t)$ is given by
\begin{eqnarray}
\Delta(\omega,t) &\equiv& \int_{0}^{t} ds
\int_{0}^{t} ds' g_{+}(t-s) \cos \omega (s-s') g_{+}(t-s').
\nonumber \\
\label{Deltaomega}
\end{eqnarray}
These results provide the tools needed to analyze the
conditions under which passage to the quantum-classical limit
may be carried out. In the next sections we present explicit
results for both the Ohmic and super-Ohmic bath spectral
densities.

\section{Thermalization of quantum brownian motion} \label{sec:QBM}

Whether a linear or nonlinear system interacting with a bath will eventually reach thermal equilibrium or not is one of the fundamental issues
in statistical mechanics. For a linear quantum brownian motion,
from a general class of initial conditions, the state of C is
considered to reach asymptotically the gaussian
form\cite{Tegmark}.
 In this section, first we will show that the
C subsystem evolution starting from an arbitrary initial condition
possesses an asymptotic limit. We then study the influence of such
a thermalized quantum system on A. For our analysis, we shall need
the results for a system composed of the particle $C$ and the
thermal bath $B$ in the absence of $A$. In this case
eq.~(\ref{Wrdensity}) takes the form,
\begin{eqnarray}
\rho_{rW}(R_t,P_t,t)&=&
 \int dR_0 dP_0 \; K_{C}(R_t, P_t;t \mid R_0, P_0; 0)\nonumber \\
&&\times \rho_{rW}(R_0,P_0,0) \;,
\label{WrdensityC}
\end{eqnarray}
In the absence of subsystem $A$, from eq.~(\ref{Ksigma}) the time
evolution kernel for the Wigner function is
\begin{eqnarray}
 K_{C}(R_t, P_t; t \mid R_0, P_0; 0) &=& N_W(t)
\exp[ - \delta \vec{X}_{W}^{T} {\bf Q}^{-1} \delta
\vec{X}_{W}], \nonumber \\
\label{Knosigma}
\end{eqnarray}
where $\vec{X}_{W}^T = (R, P)$ and
\begin{eqnarray}
\langle \vec{X} \rangle =
\left( \begin{array}{c}
        \langle R \rangle \\
        \langle P \rangle
         \end{array}      \right)
=\frac{-1}{u_{21}}
   \left( \begin{array}{cc}
       u_{11} & 1 \\
       |{\bf u}|  & u_{22}
         \end{array}      \right)
\left( \begin{array}{c}
        R_0 \\
        P_0
         \end{array}      \right).
\label{RC2}
\end{eqnarray}

For a harmonic potential, as a result of the Ehrenfest theorem,
the center phase space coordinate $\langle \vec{X} \rangle$ of
this distribution function follows a classical trajectory and will
decay because of energy dissipation into the bath.
Using the exact expressions for the matrix elements of ${\bf u}$,
obtained from the formulas given in Appendix A,
one may show that $\langle \vec{X} \rangle$ vanishes in the long
time limit for both Ohmic and super-Ohmic baths.
As a result, in both cases, for times long compared to the
characteristic relaxation time for this decay, the time evolution
kernel reduces to
\begin{eqnarray}
K_{C}(R_t, P_t; t \mid R_0, P_0; 0) &\to & N_W(t)
\exp[ - \vec{X}_{W}^{T} {\bf Q}^{-1} \vec{X}_{W} ],
\nonumber \\
&\equiv& K_{C}(R_t, P_t; t)
\label{Knosigmaasymptotic}
\end{eqnarray}
which is independent of $R_0$ and $P_0$. Consequently,
eq.~(\ref{WrdensityC}) reduces to
\begin{eqnarray}
\rho_{rW}(R_t,P_t,t)= K_{C}(R_t, P_t;t);
\label{WrdensityD}
\end{eqnarray}
thus, for an arbitrary initial condition, the Wigner transformed
reduced density matrix is uniquely determined by the time evolution
kernel $K_C$. Furthermore, the off-diagonal elements of ${\bf
Q}$ vanish in this limit and we obtain a Gaussian form
for the density matrix whose widths are uniquely specified and
given by $\langle (\Delta R)^2 \rangle=Q_{11}/2$ and
$ \langle (\Delta P)^2 \rangle=Q_{22}/2$.
The width $\langle (\Delta R)^2 (\infty) \rangle$ can be obtained from
eq.~(\ref{Q11}) as
\begin{eqnarray}
\langle (\Delta R)^2 (\infty)\rangle&=&  \frac{ \hbar}{\pi M^2}
 \int_{0}^{\infty} d \omega
 \coth \left( \frac{\beta \hbar \omega}{2} \right)
J_B(\omega) \Delta(\omega,\infty), \nonumber \\
\end{eqnarray}
We show in Appendix A for the Ohmic case and for the weak
coupling limit of the super-Ohmic case, that
$J_B(\omega) \Delta(\omega,\infty)=M^2 \chi''_C (\omega)$, where
\begin{eqnarray}
 \chi''_C (\omega) = \frac{1}{M} \frac{\omega \gamma'(\omega)}
  { (\omega_{r}^2 - \omega^2 + \omega \gamma''(\omega))^2
  + \omega^2  \gamma'^{2}(\omega) },
 \label{chi}
\end{eqnarray}
is the dynamical susceptibility for $C$.
For the spectral density defined in eq.~(\ref{SpectralDensity}),
the explicit form of the frequency dependent dissipation coefficient
$\gamma(\omega)$ is given by
\begin{eqnarray}
 M \gamma(\omega) &\rightarrow&
\left\{
\begin{array}{llll}
\eta
 \hspace{1cm} &(\nu=1)&\\
\eta \omega^2 \hspace{1cm}
&(\nu=3)&
\end{array}
\right.
\label{gammaexact2}
\end{eqnarray}
in the limit of a large cutoff parameter $\Lambda$, where
$\gamma''(\omega)$ was absorbed in the mass renormalization for the
$\nu=3$ case. Consequently, we may write
$\langle (\Delta R)^2 (\infty)\rangle$ as
\begin{eqnarray}
 \langle \Delta^2 R(\infty) \rangle
=  \frac{\hbar}{\pi}
 \int_{0}^{\infty} d \omega \coth \left( \frac{\beta \hbar \omega}{2} \right)
 \chi''_C (\omega).
\label{FDR}
\end{eqnarray}
Equation~(\ref{FDR}) is a statement of the fluctuation-dissipation
relation \cite{Kubo57}. The form in eq.~(\ref{FDR}) suggests that
the combined system $CB$ is now in thermal equilibrium with
spectrum specified by $\chi''_C (\omega)$ but the system $C$
itself is not in equilibrium. The fluctuation-dissipation relation
is violated at finite $t$ while the system is still far from
equilibrium and holds only asymptotically. 
This result generalizes the result previously 
obtained by Caldeira and Leggett
for an Ohmic bath\cite{CaldeiraLeggett84}.

When the $CB$ system is in thermal equilibrium, we show in
Appendix~\ref{app:LRT} using linear response theory that
$J_{CB}(\omega) =
\lambda^2 \chi''_C (\omega)$: the effective spectral density is
proportional to the dynamical susceptibility of the $C$ system.
If we use the relation
\begin{eqnarray}
  J_{B}(\omega)  =   M \omega \gamma'(\omega),
\label{ICBtogamma}
\end{eqnarray}
for the bath spectral density assumed in eq.~(\ref{SpectralDensity}), we see
that the effective spectral density in the combined  system $CB$ can
be written as
\begin{eqnarray}
 J_{CB}(\omega)  = \frac{\lambda^2  J_{B}(\omega) }
  { M^2 (\omega_{r}^2 - \omega^2 + \omega \gamma''(\omega))^2
  + J_{B}^2(\omega) }.
\label{EffectiveSpectralDensity}
\end{eqnarray}
This relation can also be obtained by solving the classical
equations of motion as suggested earlier \cite{FeyVer63,Leggett84,GOA85}
owing to the linear coupling assumed between the $C$ system and the bath.

\section{Emergence of quantum-classical dynamics via decoherence}
\label{sec:emergence}

The investigation of the emergence of quantum-classical dynamics
presented here will be restricted to cases where the dynamics of subsystem
$A$ occurs on time scales which are very long compared to those
that characterize subsystem $C$. We suppose that subsystem $A$ is a
two-level system with $\sigma=\pm1$ and focus on the extreme
non-adiabatic regime where the population dynamics of subsystem $A$ is
essentially frozen on the time scales of interest so that
$\rho_{A}(-1,-1,t)=\rho_{A}(-1,-1,0),\rho_{A}(1,1,t)=\rho_{A}(1,1,0)$.
We further assume that subsystems $A$ and $C$ are decoupled initially
and use the factorized initial condition,
\begin{eqnarray}
\hat{\rho}_{AC}(x_0, x'_0, 0) = \hat{\rho}_A(0) \otimes \rho_{C}(x_0, x'_0, 0).
\label{DMACinitial}
\end{eqnarray}
This is a reasonable assumption if they are weakly coupled.

\subsection{Decoherence in $A$ interacting with an equilibrium $CB$ bath}
If the interaction between the $A$ system and the combined system
$CB$ is turned on after an equilibrium state is reached, the
effect of $CB$ on $A$ is
given simply by the harmonic bath directly coupled to the $A$
subsystem but with its spectrum characterized by the effective spectral
density. Under these conditions, the off-diagonal part of the density matrix
can be obtained from eq.~(\ref{DMACf}) by dropping the dependence on
$(x,x')$, omitting the integrals over spin variables and by making the
variable replacements $x \rightarrow \sigma$ and $J_B \rightarrow J_{CB}$
in eq.~(\ref{IFaction}) for the influence functional.  We find,
\begin{eqnarray}
  \rho_A(-1,1, t)&=&   e^{-D_{A}^{(0)}(t)}   \rho_A(-1,1, 0),
\label{ODDM}
\end{eqnarray}
where $D_{A}^{(0)}(t)$ is a decoherence factor for A and is defined as
\begin{eqnarray}
D_{A}^{(0)}(t) &=& \frac{4}{\pi \hbar}
 \int_{0}^{\infty} d \omega  J_{CB}(\omega)
 \coth \left( \frac{\beta \hbar \omega}{2} \right)
\nonumber \\ &&\times \left(   \frac{1 - \cos(\omega t)}{\omega^2}  \right)\;.
\label{ODDM2}
\end{eqnarray}
Thus, as mentioned above, the dynamics of the composite system is equivalent
to that of a two-level spin-boson system coupled to a harmonic bath,
with an effective spectral density.

\begin{figure}[htbp]
\epsfxsize=.45\textwidth \epsfbox{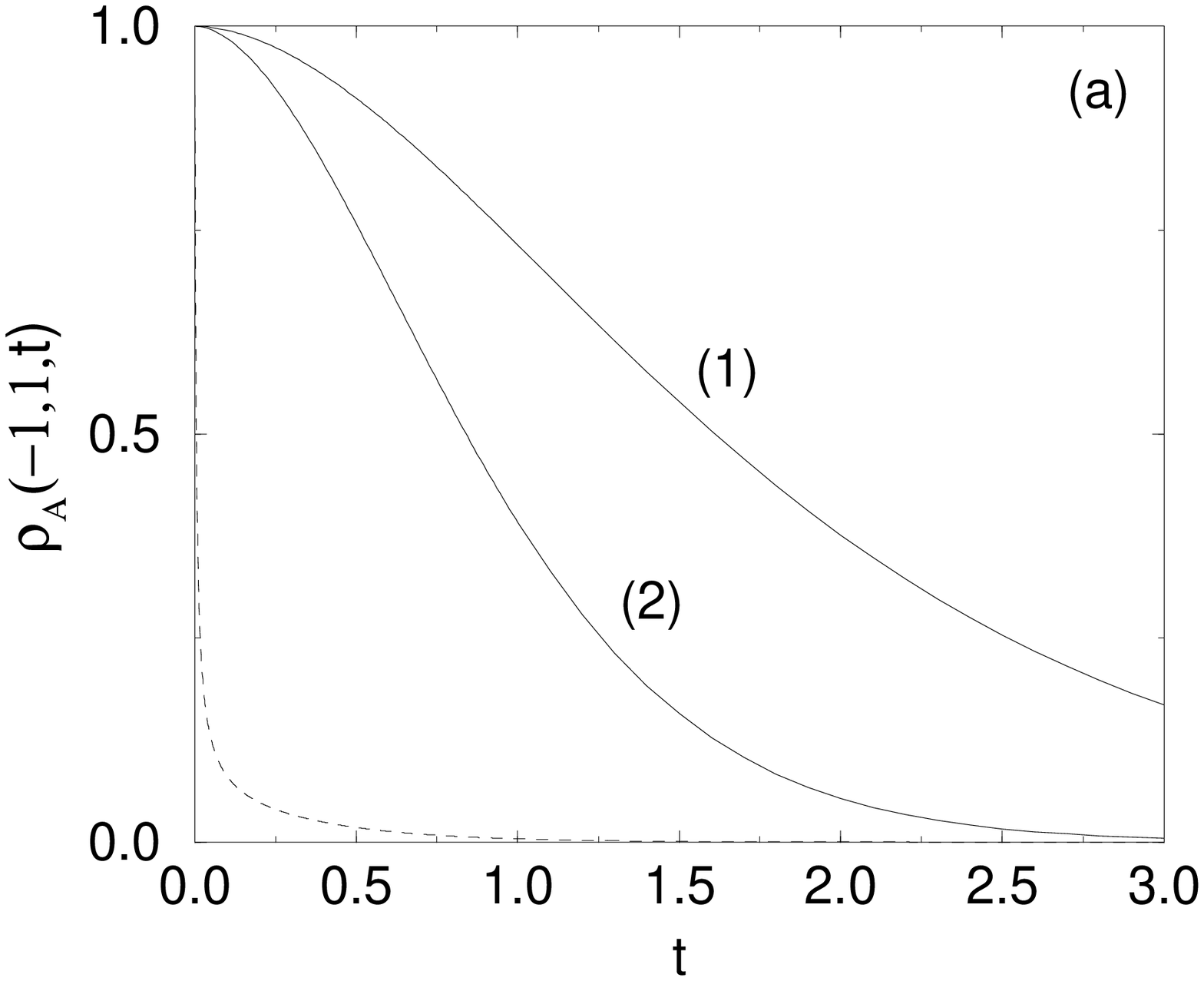}
\epsfxsize=.45\textwidth \epsfbox{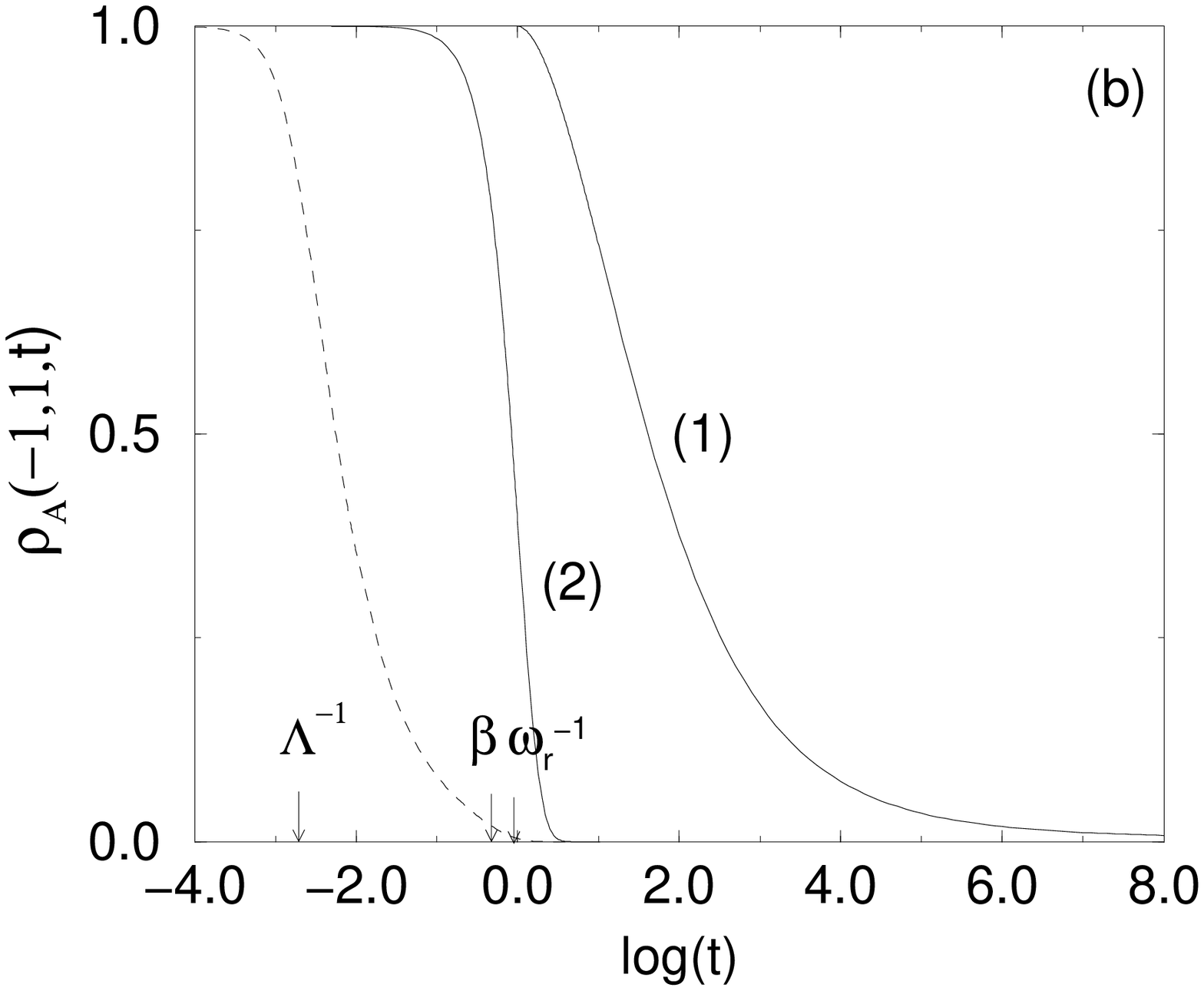}
\caption{ The temporal behavior of the off-diagonal element of the
reduced density matrix $\rho_A(-1,1,t)$ for subsystem $A$ is plotted
in panel (a) versus time. The initial value $\rho_A(-1,1,0)=1$. Parameters are:
$\beta=0.66$, $\Lambda=500$, $\omega_{r}=1$, $\gamma=0.3$,
$M=1$ and $\hbar=1$.  The same quantity is shown in panel (b) using a logarithmic
time scale. In both figures the solid line 
corresponds to $A$ under the
influence of $CB$
with the effective Ohmic spectral density,
while the dashed line shows the influence of Ohmic bath $B$ without $C$.
(1) $\lambda^2=0.1$ and (2) $\lambda^2=0.3$.}
\label{fig1}
\end{figure}
Using these results we may examine the coherence in $A$ under the influence
of the effective bath $CB$ and the coherence of $A$ under the influence of
the bath $B$ separately. In Figs.~1(a) and (b), the off-diagonal part
of the density matrix of subsystem $A$ is plotted for an Ohmic environment.

The solid line corresponds to our system while the dashed line
is the ordinary spin boson model without $C$.
The time evolution of $\rho_A(-1,1, t)$ for the ordinary ordinary spin
boson model is characterized by (1) an initial
period where subsystem $A$ system has not yet felt the
existence of the environment and its coherence is maintained; (2)
a quantum regime for $t> 1/\Lambda$ where the system begins to interact
with the vacuum fluctuations of the environment and, finally, (3) a thermal
regime for $t> \hbar \beta$ where the effects of thermal fluctuations
on subsystem $A$ have set in. For $1/\Lambda < t < \hbar \beta$,
only vacuum fluctuations interact with the system.

By contrast, the existence of the intermediate subsystem $C$ changes
the evolution of $\rho_A(-1,1, t)$ significantly. Subsystem $C$ is
characterized by its harmonic oscillator frequency and mass. Fluctuations
originating in the bath $B$ are modulated by $C$ through these parameters.
For a large mass $M$, the effective modes of the combined $CB$ system are
concentrated in the neighborhood of $\omega_{r}$ and the decoherence
behavior of subsystem $A$ is governed by these modes.
Therefore, after the initial period, decoherence will begin to occur at
$t \sim 1/ \omega_r$.

\begin{figure}[htbp]
\epsfxsize=.45\textwidth \epsfbox{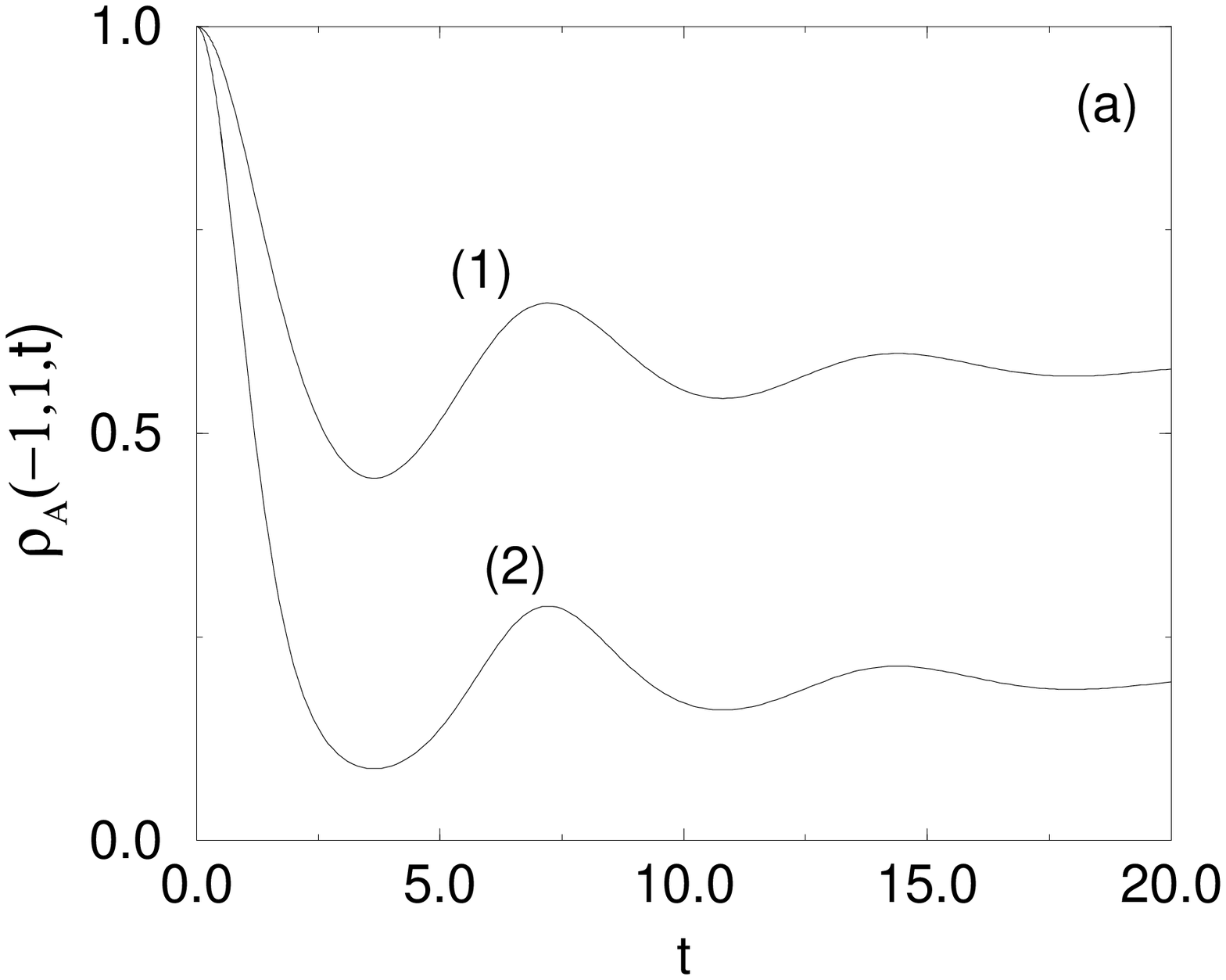}
\epsfxsize=.45\textwidth \epsfbox{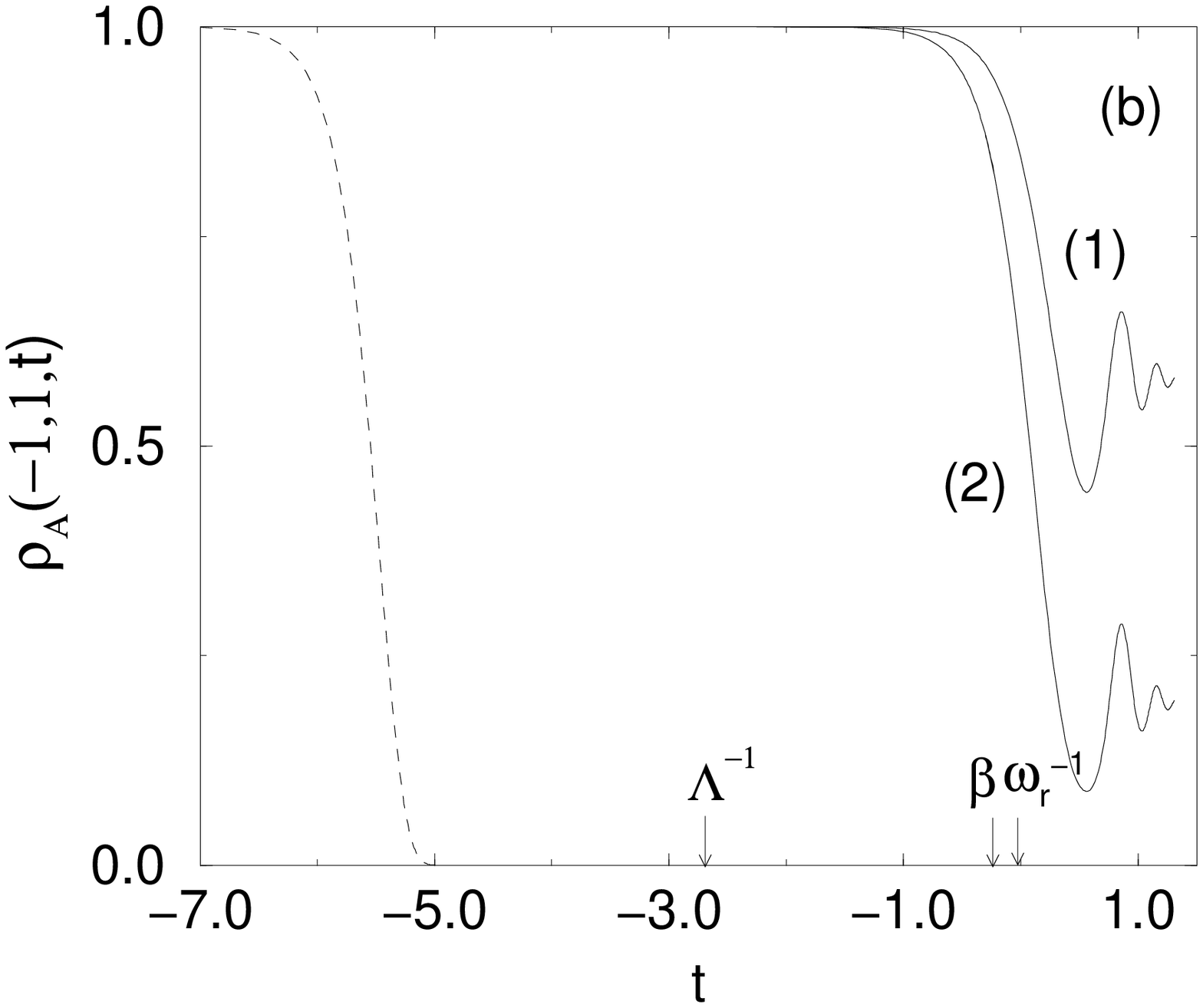}
\caption{ Plot of $\rho_A(-1,1,t)$ versus time (a) and versus $\log(t)$ (b)
for a super-Ohmic environment with $\nu=3$. Parameters  and labeling
are the same as in \protect Fig.~\ref{fig1}. The dashed line in panel
(a) for super-Ohmic bath $B$ without $C$ overlaps the vertical axis.
The behavior is visible in panel (b) using the logarithmic time scale. }
\label{fig2}
\end{figure}

Although some systems are approximately characterized by an Ohmic
spectral density, generic systems will have non-Ohmic spectral densities.
The density of states typically varies as $\omega^{\nu}$ with
$\nu$ depending on the spatial dimension $D$. We consider the 3D
super-Ohmic case with $\nu=3$. Such an environment is
relevant for the study of polarons\cite{LCDFGZ87},
macroscopic magnetization tunneling in crystals\cite{GargKim89},
and radiation damping of atoms
\cite{BaroneCaldeira91}. A super-Ohmic environment
affects subsystem $A$ evolution on short time scales more significantly
than an Ohmic environment. A sub-Ohmic environment induces non-trivial
long time behavior. Results for the super-Ohmic case with $\nu=3$ are
shown in Figs.~\ref{fig2} (a) and (b).

For the super-Ohmic high dimensional environment, the vacuum fluctuations
play a more significant role than for an Ohmic environment, reflecting the
fact that there are larger number of high frequency modes in the higher
dimensional environment. The same effect is also responsible
for the large difference in the behavior of
correlation functions of the system in quantum and classical baths
when the bath is super-Ohmic\cite{EgoEveSki99}.
The dynamics of $A$ in the presence of the intermediate subsystem
$C$ is worth noting. For the present choice of parameters,
$\rho_A(-1,1, t)$ asymptotically reaches a non-zero value indicating
that quantum coherence in the $A$ system will never be lost completely.
The origin of this behavior can be understood 
in terms of the time-dependence of the diffusion coefficient 
(See Ref.~\cite{QBM1}).
System $C$ initially executes brownian motion with
a time-dependent diffusion coefficient
as a result of the non-Markovian time evolution of the reduced density matrix.
This time-dependent diffusion coefficient of $C$, which also determines
the decoherence rate,
exhibits a rapid increase
at early times $t > 1/\Lambda$ and then asymptotically vanishes.
This initial increase can be large enough to wash away the
quantum coherence of $C$ directly coupled to the bath.
Owing to the modulation effect from $C$, these high frequency modes are
filtered out and do not directly affect $A$ if $\omega_{r}<<\Lambda$.
The late time value of the diffusion coefficient is too small to eliminate
the quantum coherence from subsystem $A$ completely.
Hence, the non-Markonvian nature of the density matrix evolution 
for the super-Ohmic case is responsible for the significant difference
in the dynamics of coherence between the $A$ and $C$ subsystems.

\subsection{Decoherence in $A$ for arbitrary $C$ initial conditions}

Thus far we have considered only the asymptotic limit where $CB$ has
reached equilibrium and the fluctuation-dissipation relation holds.
It is interesting to consider the coherence of $A$ under the
influence of intrinsic dynamics of $C$ starting from arbitrary $C$
initial conditions. As an example of this situation, we assume that $C$
is initially in a gaussian coherent state given by the wave function,
\begin{eqnarray}
 \psi_{\bar{x} \bar{p}}(x) &=& \frac{1}{(\pi \epsilon^2)^{1/4}} \exp \left[
 -\frac{(x-\bar{x})^2}{2 \epsilon^2} + i \frac{\bar{p} x}{\hbar}  \right],
\label{WF0}
\end{eqnarray}
with width $\epsilon=1/\sqrt{\omega_r}$ where $\bar{x}$ and $\bar{p}$
are parameters. The $C$ subsystem density matrix at the initial time is then
given by
\begin{eqnarray}
 \rho_{C}(x_0, x'_0, 0) &=&\frac{1}{\sqrt{\pi \epsilon^2}}
\exp \left[  -\frac{(R_0-\bar{x})^2}{\epsilon^2} -\frac{r_0^2}{4 \epsilon^2}
 +i \frac{\bar{p} r_0}{\hbar}  \right].
 \nonumber \\
\label{2IDMx}
\end{eqnarray}
We may now use this expression for $\rho_{C}(x_0, x'_0, 0)$ in the
factorized initial condition for $AC$ given in eq.~(\ref{DMACinitial}).
The off-diagonal element of reduced density matrix for subsystem $A$
in the non-adiabatic limit can then be determined from
eqs.~(\ref{DMACf}) and (\ref{Jrpathint}) for
$\rho_r(\sigma, \sigma', x, x', t)$ for $AC$ by integrating over the
$C$ variables. The result is,
\begin{eqnarray}
\rho_A(-1,1,t)&=&\int dx dx'\; \Big[\int dx_0 dx'_0\; \nonumber \\
&& \times J^{-1,1}_C(x,x',t|x_0,x'_0,0)\rho_{C}(x_0, x'_0, 0)\Big] \nonumber \\
&& \times \rho_A(-1,1,0)\;.
\end{eqnarray}
Changing integration variables $(x,x') \to (R,r)$
and $(x_0,x_0') \to (R_0,r_0)$ and using eq.~(\ref{Jrsimple}) for
$J^{-1,1}_C(R,r,t|R_0,r_0,0)$ we may write this as
\begin{eqnarray}
\rho_A(-1,1,t)&=&\int dR dr\; \Big[\int dR_0 dr_0\; \nonumber \\
&& \times N(t) \exp\Big\{\frac{i}{\hbar} {\mathcal L} \Big\}
\rho_{C}(R_0, r_0, 0)\Big] \nonumber \\
&& \times \rho_A(-1,1,0)\;.
\end{eqnarray}
Making use of the expression for ${\mathcal L}$ evaluated for
$\sigma=-1$ and $\sigma'=1$ and carrying out the integrations we find,
\begin{eqnarray}
 \rho_A(-1,1, t)&=&  \rho_A(-1,1, 0)   e^{-D_A(t)} \nonumber \\
&& \times
  \exp \left[\frac{i}{\hbar} ( f_{R}(t) \bar{x} + f_{P}(t) \bar{p} ) \right]
\nonumber \\
&& \times
\exp \left[-\frac{1}{\hbar} ( f_{R}^2(t)/ M \omega_r +  f_{P}^2(t) M \omega_r )
  \right]\:, \nonumber \\
 \label{DMAnonadiabatic_offdiag}
\end{eqnarray}
where
\begin{eqnarray}
 f_{R}(t)  = \lambda  \int_{0}^{t} ds  \frac{u_{11}(s)}{u_{21}(s)},
 \quad  f_{P}(t)  = \lambda  \int_{0}^{t} ds  \frac{1}{u_{21}(s)}
\label{AB}
\end{eqnarray}
and
\begin{eqnarray}
 D_A(t) =  \frac{2 \lambda^2}{\pi \hbar M^2}
 \int_{0}^{\infty} d \omega  J_{B}(\omega)
 \coth \left( \frac{\beta \hbar \omega}{2} \right)  \Delta_A(\omega,t).
 \nonumber \\
\label{Decfunctionfull}
\end{eqnarray}
The explicit expression for $\Delta_A(\omega,t)$ can be computed
from the solutions presented in Appendix~A by a lengthy but
straightforward calculation. We find,
\begin{eqnarray}
&&\Delta_A(\omega,t) = \frac{1}{(\Omega_{r}^2 - \omega^2)^2 + 4
\Gamma^2 \omega^2 } \left[ 2 \frac{1 - \cos(\omega t)}{\omega^2}
\right. \nonumber \\
&& \quad +\frac{2}{\Omega \Omega_{r}^2} \frac{1 -
\cos(\omega t)}{\omega}\nonumber \\
 & &\quad \times
\left( \omega \Omega (1-e^{-\Gamma t} \cos(\Omega t) )
 - \omega \Gamma    e^{-\Gamma t} \sin(\Omega t)\right)\nonumber \\
&&\quad+ \frac{2}{\Omega \Omega_{r}^2} \frac{\sin(\omega
t)}{\omega}
\nonumber \\
& &\quad \times \left( 2 \Gamma \Omega (1-e^{-\Gamma t}
\cos(\Omega t) ) + (\Omega^2 - \Gamma^2) e^{-\Gamma t} \sin(\Omega
t) \right)
\nonumber \\
&& \quad +\frac{1}{\Omega^2 \Omega_{r}^{4}} \left\{  \left(
\omega^2 + 4 \Gamma^2 \right) \Omega^2 ( 1- \cos(\Omega t) e^{-
\Gamma t} )^2
\right. \nonumber \\
&& \quad -
 2 \Gamma \Omega   \left( \omega^2 + 2 \Gamma^2 - 2 \Omega^2 \right)
\nonumber \\
&& \quad \times ( 1 - \cos(\Omega t) e^{- \Gamma t} )\sin(\Omega
t)e^{- \Gamma
t}\nonumber \\
&& \quad +\left. \left.
 \left( \omega^2 \Gamma^2 + (\Omega^2 - \Gamma^2)^2 \right)
 \sin^2(\Omega t) e^{- 2 \Gamma t}\right\}\right].
\label{DeltaAfullOhmic}
\end{eqnarray}
Equations (\ref{DMAnonadiabatic_offdiag})-(\ref{DeltaAfullOhmic}) are
our main results. The first term in
eq.~(\ref{DMAnonadiabatic_offdiag}) is the initial condition for
$A$, the second term accounts for the decoherence arising from
thermal fluctuations of the bath mediated by $C$. The third term
arises from the initial condition for $C$ and the last term is
responsible for the decoherence of $A$ due to the averaged damped
oscillatory motion of $C$.

\begin{figure}[pp]
\epsfxsize=.45\textwidth \epsfbox{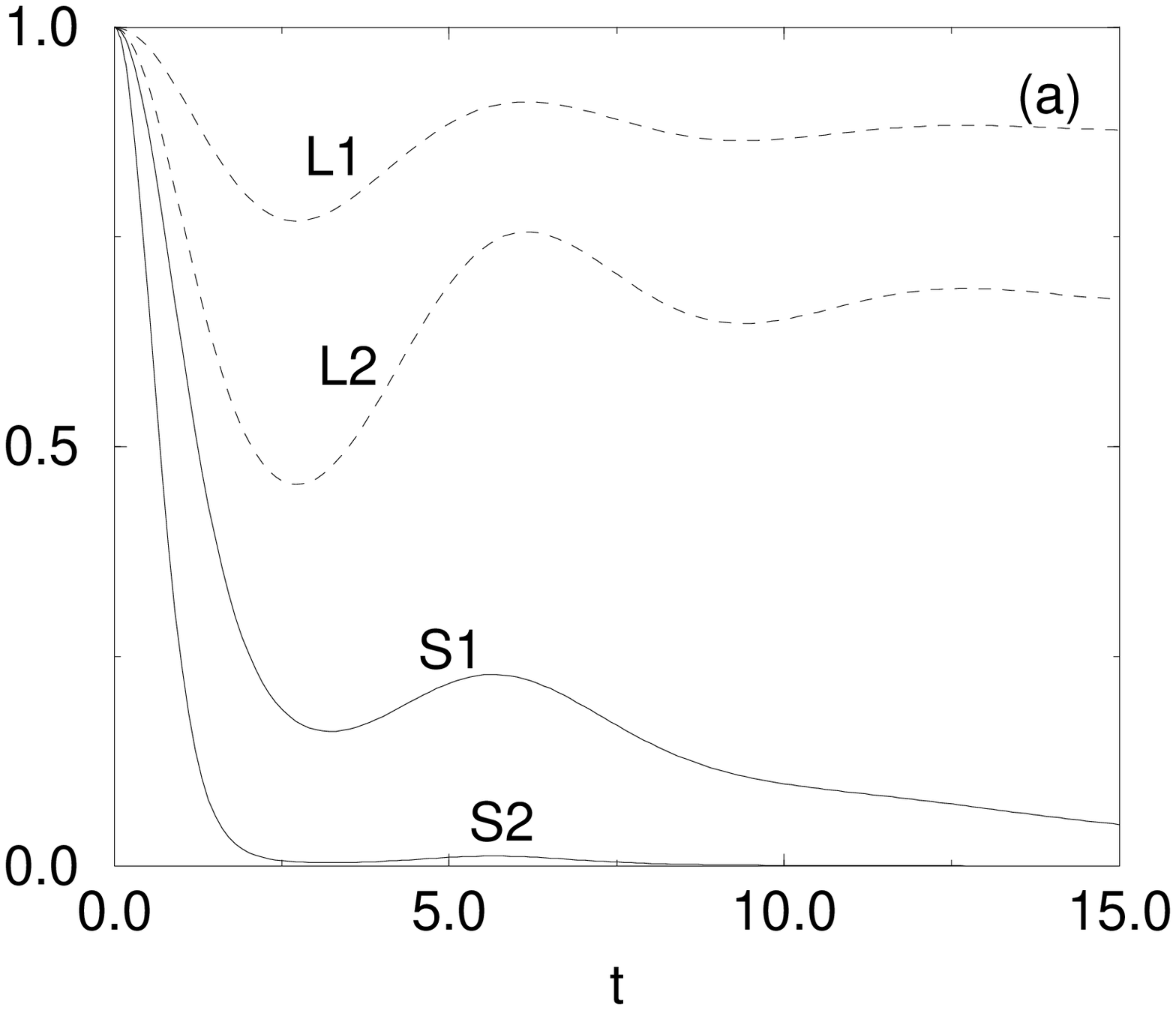}
\epsfxsize=.45\textwidth \epsfbox{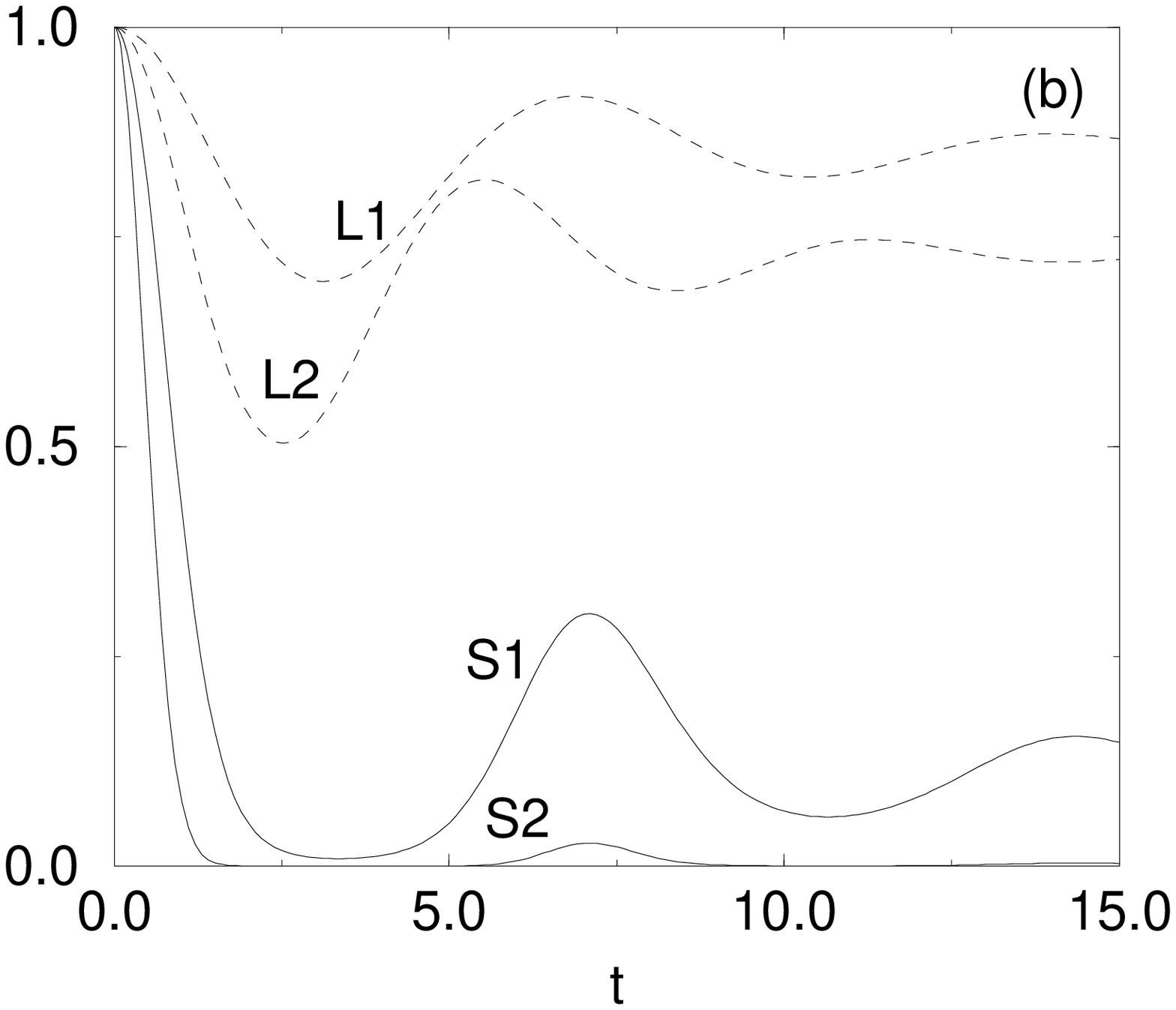}
\caption{
Plot of contribution to the modulus of $\rho_A(-1,1,t)$ versus time for an Ohmic bath
(Fig.~\ref{fig3}(a)) and for a super Ohmic bath (Fig.~\ref{fig3}(b)).
Subsystem $C$ is in a coherent state initially.
In both figures the solid line corresponds to the second term S,
while the dashed line corresponds to the last term L 
in the right hand side in eq.(\ref{DMAnonadiabatic_offdiag}).
$\lambda^2=0.1$ for S1,L1 and $\lambda^2=0.3$ for S2,L2.
Parameters values are the same as in \protect Fig.~\ref{fig1}.
}
\label{fig3}
\end{figure}

The modulus of the off-diagonal reduced density matrix element for $A$
is shown in Figs.~\ref{fig3}(a) and (b) as a function of time.
The solid line shows the second term eq.~(\ref{DMAnonadiabatic_offdiag}),
$S \equiv e^{-D_A(t)}$,
which gives the decoherence due to the quantum-back-reaction-induced
self-interaction of subsystem $A$. 
The dashed line shows the last term, 
$L \equiv \exp \left[-\frac{1}{\hbar} ( f_{R}^2(t)/ M \omega_r +  f_{P}^2(t) M \omega_r) \right]$,
the decoherence
due to the motion of $C$. The latter effect
is not strong enough to eliminate coherence of $A$ completely.
Since, initially, the CB is far from equilibrium,
those modes that affect A are not necessarily
near resonant modes around $\omega_r$.
Consequently, the decoherence of A is more rapid
in this case than in the case of evolution 
from the thermal equilibrium initial condition
studied in Sec. IV. A.
This tendency is more evident in a super-Ohmic bath than in an Ohmic bath.
With generic initial conditions for $C$,
A is under the influence of the nonequilibrium bath $CB$,
which is no longer equivalent to the effective thermal bath with the 
spectral density $J_{CB}$. 

Now let us consider the situation where the decoherence of $C$ is
fast and that of $A$ is slow. This may occur at high temperatures
and for weak coupling between $A$ and $C$. In such a case, the
back reaction from $A$ on $C$ may be neglected. In this
circumstance, we may consider the decoherence of $C$ in the
absence of $A$. Quantum brownian motion and decoherence of a
damped harmonic oscillator has been studied previously
\cite{CaldeiraLeggett84,GSI88,UnruhZurek89,QBM1,QBM2}. Using the
result in eq.~(3.9) of Ref.~\cite{QBM1}, the reduced density matrix may
be approximated by,
\begin{eqnarray}
 \rho_{r}(\sigma,\sigma',x,x',t) \sim e^{-(x-x')^2 D_C(t)}
 \rho_{r}(\sigma,\sigma',x,x',0),
\label{MasterEquationDecC}
\end{eqnarray}
where $D_C(t) =\int_{0}^{t} ds \int_{0}^{s} ds'\alpha^R(s') \cos
\omega_r s'$ is the decoherence factor for $C$.
\begin{figure}[hh]
 \begin{center}
\epsfxsize=.45\textwidth \epsfbox{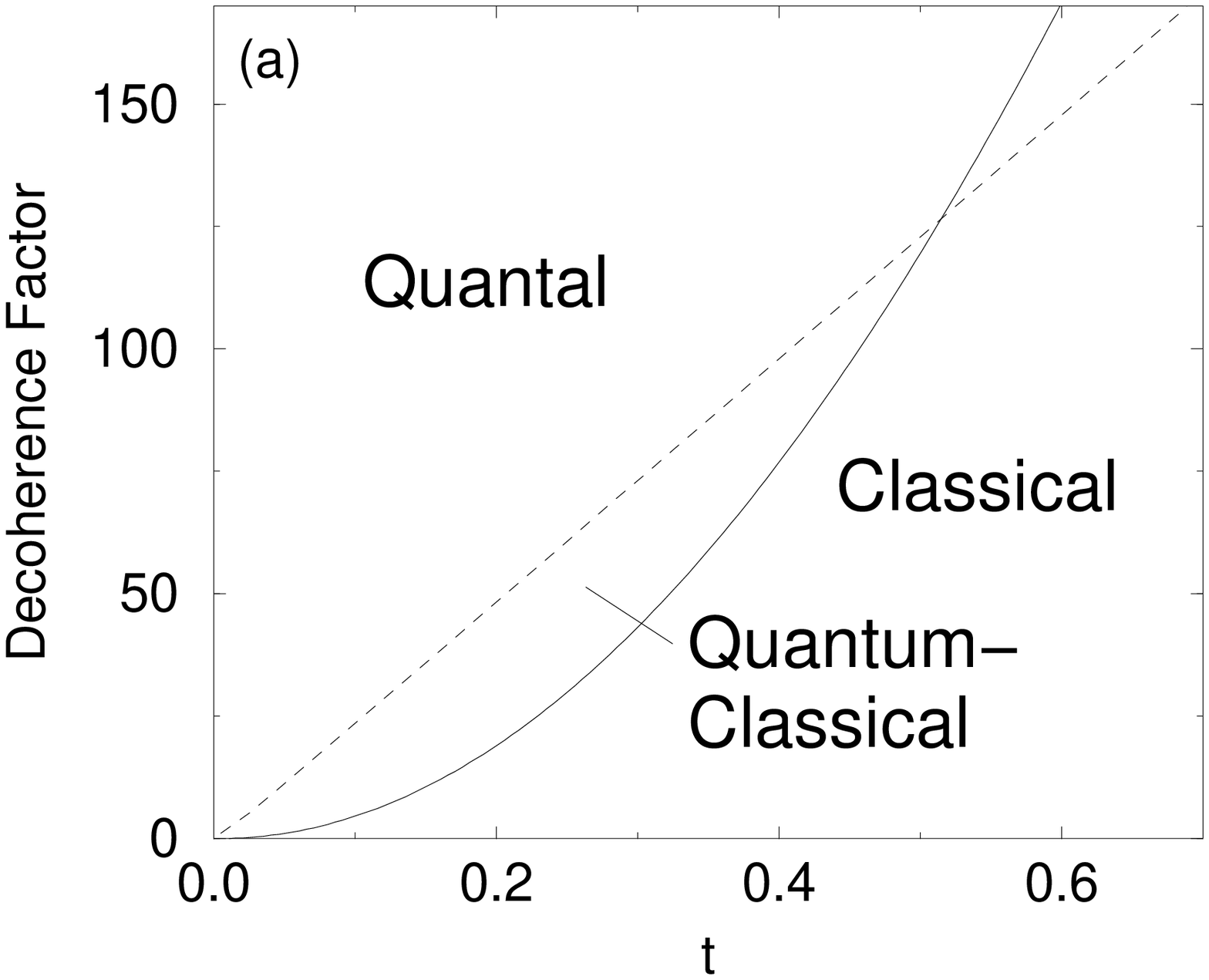}
\epsfxsize=.45\textwidth \epsfbox{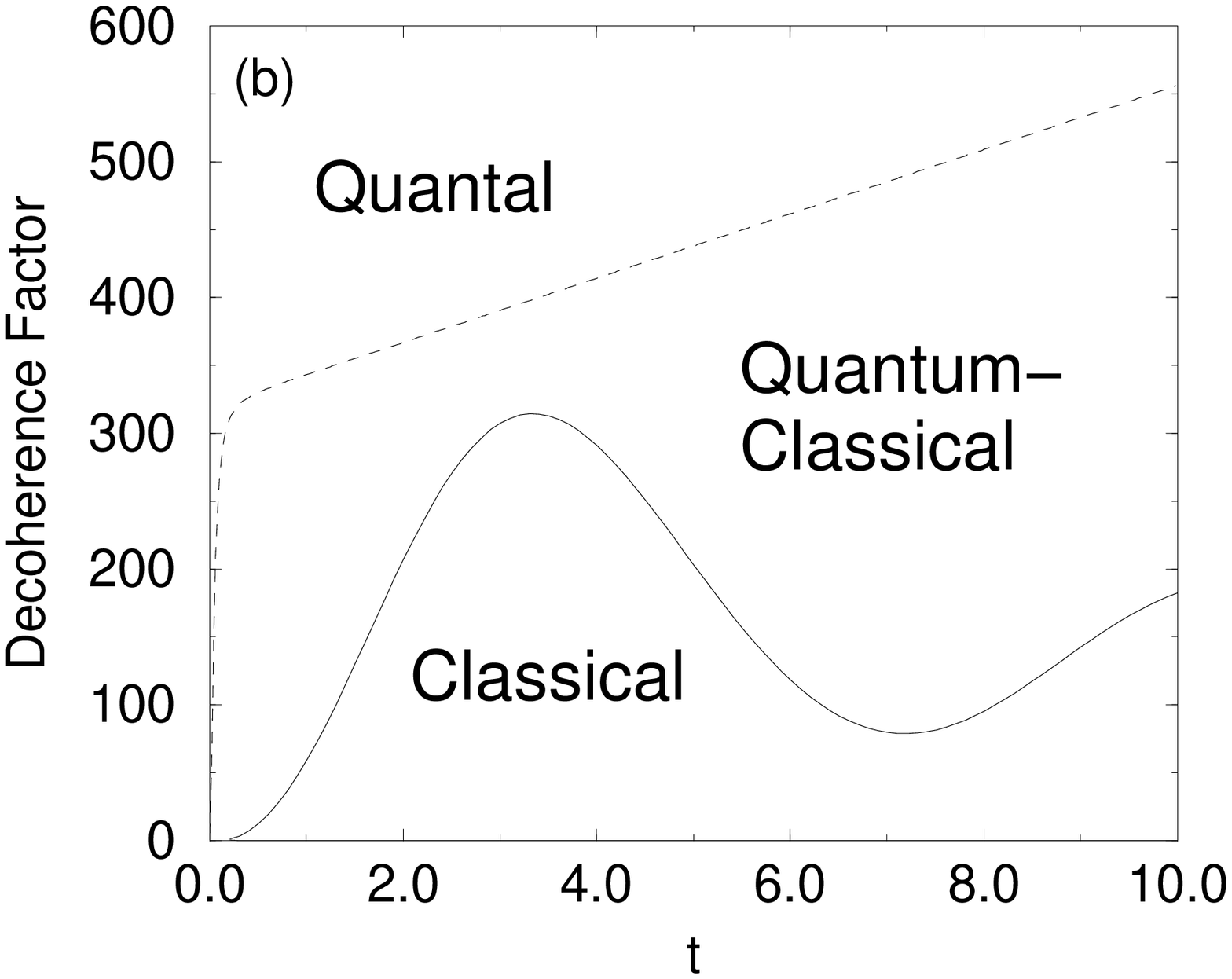}
  \end{center}
\caption{
Plot of the decoherence factor versus time. The solid line corresponds
to the logarithm of the modulus of
the density matrix $\rho_A$ in eq.(\ref{DMAnonadiabatic_offdiag})
with the inverse sign,
while the dashed line is $D_C$.
System parameters are:
$x-x'=1$, $\omega_{r}=1$, $\gamma=0.3$, $M=1$, $\hbar=1$.
Panel (a) is for the Ohmic bath with $\beta =0.002$, $\Lambda=500$.
Panel (b) is for the $\nu=3$ super-Ohmic bath with $\beta =0.02$,
$\Lambda=20$.}
\label{fig4}
\end{figure}
In Fig.~\ref{fig4}, the
decoherence factors for $A$ and $C$ are plotted.
For an Ohmic bath (Fig.~\ref{fig4}(a)), initially,
the quantum coherence is
lost faster $C$ than in $A$; hence, the $A$ subsystem behaves more
quantum mechanically than the $C$ subsystem during this initial
period. Figure~\ref{fig4}(a) shows that there is a crossover and
for longer times, $t>1/\omega_r$,  subsystem $A$ experiences
stronger decoherence than $C$. The decoherence factor for $C$
varies linearly with $t$ indicating that the dynamics of $C$
asymptotically approaches Markovian evolution in the high temperature
Ohmic bath.
For a super-Ohmic bath (Fig.~\ref{fig4}(b)), a substantial portion
of the quantum coherence in $C$ is lost during the initial period
due to large fluctuations coming from the bath, while $A$ retains
quantum coherence for a long time.  The dynamics of the
coupled system can be approximated by mixed quantum-classical
evolution for long times after the short transient period.

\section{Conclusion} \label{sec:conc}

An appreciation of the conditions under which a quantum mechanical
system may be approximated as a quantum-classical system is
essential in many applications to the dynamics of many-body
systems. Adopting an open quantum system point of view has
provided a natural way to explore this issue based on the
decoherence of the superposition of quantum states into a
statistical mixture under the influence of the environment
\cite{EID,Dec96}. Indeed, considerations of decoherence have played an
important role in discussions of schemes for the simulation of
condensed phase systems. \cite{decoh}

Using influence functional methods, we have shown for our simple
model system that the decoherence time scales that characterize
the $A$ and $C$ subsystems can differ significantly. In
particular, in the limit of nonadiabatic dynamics we have
identified the following three regimes: (1) the full quantum
regime where both the $A$ and $C$ subsystems behave quantum
mechanically; (2) the quantum-classical regime where
subsystem $A$ maintains coherence owing to its indirect coupling
to the bath, while $C$ has lost its coherence and behaves
effectively classically; (3) the classical regime where the
quantum coherence of both the $A$ and $C$ subsystems has been lost
and the composite $AC$ subsystem  exhibits effectively classical
dynamics. 
The different roles that the environment plays
in its effect on two subsystems is responsible for the separation
of decoherence time scales.

We also saw that a different choice of initial conditions will
modify above picture. When the initial condition for $C$ is chosen
to be a pure gaussian coherent state, $CB$ can no longer be
considered to be a thermal bath and the coherence of $A$ will be
lost in a shorter time scale 
than times given by the inverse characteristic
frequency of $C$.

Quantum/classical correspondence in nonlinear systems, in
particular, in chaotic systems, has many nontrivial
features\cite{Chi}. The quantum open system approach adopted in this
paper can also shed light on this problem\cite{ShiokawaHu95,PazZurek00}.
While the model system studied here is very simple and the extreme
non-adiabatic regime is only one limiting case to consider, our
results nevertheless have provided some insight into the emergence
of quantum-classical dynamics and should be useful in the study of
the quantum dynamics of more complex systems.

\section*{Acknowledgments}
This  work  was  supported in part by a grant from the Natural
Sciences and Engineering Research Council of Canada.
Acknowledgment is made to the donors of The Petroleum Research
Fund, administered by the ACS, for partial support of this
research.

\appendix
\renewcommand{\theequation}{\thesection\arabic{equation}}
\setcounter{equation}{0}

\section{Solutions of equations of motion} \label{app:Formulas}

In this Appendix we give some details of the solutions of the
equations of motion discussed in Sec.~\ref{EL} which are used in
the calculations presented in Sec.~\ref{sec:emergence}. We require
the functions $u_{1,2}(s)$ (and $v_{1,2}(s)$) which are the
solutions of the homogeneous parts of the Euler-Lagrange equations
(\ref{EL1}) and (\ref{EL2}). These functions can be found if
$g_{\pm}(s)$, which are also solutions of the homogeneous parts
eqs.~(\ref{EL1}) and (\ref{EL2}), are known, since
\begin{eqnarray}
u_{1}(s)  = \dot{g}_{+}(s) - \frac{g_{+}(s)}{g_{+}(t)} \dot{g}_{+}(t),
\qquad u_{2}(s) = \frac{g_{+}(s)}{g_{+}(t)},
\label{u1u2}
\end{eqnarray}
in order to satisfy the boundary conditions on these functions. Using
the spectral density for an Ohmic bath, the equation that $g_{\pm}(s)$
satisfies is
\begin{eqnarray}
\ddot{g}_{\pm}(s) \pm 2 \Gamma \dot{g}_{\pm}(s) + \omega_{r}^2
g_{\pm}(s)=0\;. \label{EOMg}
\end{eqnarray}
The solution of eq.~(\ref{EOMg}) is
\begin{eqnarray}
g_{\pm}(s)  &=&  \frac{\sin \Omega s}{\Omega } e^{\mp \Gamma s}\;,
\label{g+-exp}
\end{eqnarray}
where $\Gamma=\gamma$ and $\Omega^2 \equiv \omega_{r}^2
-\gamma^2$.

 Using the spectral density for a super-Ohmic bath with
$\nu=3$, the equations of motion for $g_{\pm}(s)$ can be written
as
\begin{equation}
\ddot{g}_{\pm}(s)  + \omega_{r}^2 g_{\pm}(s) \mp 2\gamma
\stackrel{...}{{g}}_{\pm}(s)=0\;. \label{AL}
\end{equation}
Thus, the equation of motion has the form of the Abraham-Lorentz
equation \cite{Jackson75}. This equation is an approximation to
the full equation of motion with non-local time dependence. As in
the electromagnetic field case, only the physically relevant roots
of the characteristic equation must be retained. These are $\Gamma
\pm i\Omega$ with
\begin{eqnarray}
\Gamma \equiv \frac{1}{6 \gamma} \left(\frac{D}{4}+\frac{1}{D} -1
\right), \qquad \Omega \equiv \frac{\sqrt{3}}{6 \gamma}
\left(\frac{D}{4}-\frac{1}{D} \right) \label{LambdaGammaOmega}
\end{eqnarray}
where
\begin{eqnarray}
D = 2  \left( 1+ 54 (\gamma \omega_{r})^{2}+ \sqrt{108}\gamma
\omega_{r} \sqrt{27 (\gamma \omega_{r})^{2} +1} \right)^{1/3}.
\label{Ddef}
\end{eqnarray}
Retaining the physically relevant roots of the characteristic
equation, the solution for $g_{\pm}(s)$ for the super-Ohmic case
has the same form as eq.~(\ref{g+-exp}) but with the above values
of $\Gamma$ and $\Omega$. In the weak coupling limit that we
consider in our calculations, $\Gamma \to \gamma \omega_r^2$ and
$\Omega \to \omega_r^2$.

In our model, eq.~(\ref{AL}) can be reduced to a second order ordinary
differential equation by differentiating the homogeneous part of the
Euler-Lagrange equations with respect to time as
\begin{equation}
\stackrel{...}{{g}}_{\pm}(s)=- \omega^2_r \dot{g}_{\pm}(s)
+{\mathcal O}(\gamma)\;,
\end{equation}
which may be substituted into eq.~(\ref{AL}) to give (\ref{EOMg}) with
$\Gamma=\gamma \omega^2_r$ and
$\Omega=\omega_r$ in the leading order approximation as indicated above.
Although one can improve the
approximation leading to eq.~(\ref{EOMg}) for the super-Ohmic case
to arbitrary higher order in $\gamma$, we restrict our study to the leading
order in this parameter.

Since the $u_{1,2}$ and $v_{1,2}$ solutions are now known, we have
all the information needed to compute the various quantities
necessary to obtain the numerical results. In particular, the
expression for the matrix elements of ${\bf a}$ in eq.~(\ref{aij})
for a general environment can be found.
(See Refs.~\cite{CaldeiraLeggett84,QBM1,QBM2} for an Ohmic bath.) In our
computations, we need $a_{11}(t)$ which has the form,
\begin{eqnarray}
a_{11}(t) &=& \frac{1}{2 \pi g_{+}^2(t)} \int_{0}^{\infty} d
\omega J_B(\omega) \coth \frac{\beta \hbar \omega}{2}
\Delta(\omega,t)\;,
\label{a11general}
\end{eqnarray}
where $\Delta(\omega,t)$ was defined in eq.~(\ref{Deltaomega}).
The evaluation of $\Delta(\omega,t)$ is straightforward and leads to,
\begin{eqnarray}
&&\Delta(\omega,t)= \frac{1}{2 \Omega^2}
 \frac{1}{(\Omega_{r}^2 - \omega^2)^2 + 4 \Gamma^2 \omega^2 }
\left[    2 \Omega^2 +    e^{-\Gamma t} \right.\nonumber \\
&& \quad \times \left.
   \left[  2 \cos(\omega t)   \{ \left( \Gamma^2 + \omega_{+} \omega_{-} \right)
    \cos(\Omega t) - 2 \Gamma \Omega \sin(\Omega t)   \}   \right. \right.
   \nonumber \\
   &&\quad -
   \left. \left.
   \left( \Gamma^2 + \omega_{-}^2 \right)    \cos(\omega_{+} t)
   -\left( \Gamma^2 + \omega_{+}^2 \right)    \cos(\omega_{-} t)
   \right]\right. \nonumber \\
   &&\quad + \left. e^{-2 \Gamma t}  \{      \Omega_{r}^2 + \omega^2
      +2 \Gamma \Omega \sin(2 \Omega t) \right. \nonumber \\
&&\quad \left.
     -\left( \Gamma^2 + \omega_{+} \omega_{-} \right)  \cos(2 \Omega t)   \}
 \right]
\label{G2Ohmic}
\end{eqnarray}
where $\omega_{\pm} \equiv \omega \pm \Omega$ and $\Omega_{r}^2
\equiv \Omega^2 + \Gamma^2$. Using this result, in the asymptotic
limit we obtain,
\begin{eqnarray}
J_B(\omega) \Delta(\omega,\infty)&=&
\left\{
\begin{array}{llll}
\frac{2M \gamma \omega}{(\Omega_{r}^2 - \omega^2)^2 + 4 \gamma^2 \omega^2 }
  &(\nu=1)& \nonumber
\\
\frac{2M \gamma \omega^3}{(\Omega_{r}^2 - \omega^2)^2 + 4 \gamma^2 \omega^6 }
  &(\nu=3)&.
\end{array}
\right.
\label{G2longtime}
\end{eqnarray}
In writing this equation, we used the fact that the distribution
$J_B(\omega) \Delta(\omega,\infty)$ is highly peaked around
$\omega_r$ for small $\gamma$.

Similarly, for long times we may write the elements of ${\bf a}$
as,
\begin{eqnarray}
a_{11}(t)
&\rightarrow&
\frac{\Omega^2 e^{2 \Gamma t}}{2 \pi \sin^2 \Omega t}
\int_{0}^{\infty} d \omega J_B(\omega)
\coth \frac{\beta \hbar \omega}{2} \Delta(\omega,\infty)
\nonumber \\
a_{12}(t)  &=& a_{21}(t) \rightarrow
\frac{\Omega e^{\Gamma t}}
{2 \pi \sin \Omega t}
\int_{0}^{\infty} d \omega J_B(\omega)
\coth \frac{\beta \hbar \omega}{2}
\nonumber \\
(&\Gamma& -\Omega \cot (\Omega s) )
\Delta(\omega,\infty)
\nonumber \\
 a_{22}(t) &\rightarrow& \frac{1}{2 \pi} \int_{0}^{\infty} d
\omega J_B(\omega) \coth \frac{\beta \hbar \omega}{2} \nonumber \\
 \left\{
\right. &\omega^2& \left. + (\Gamma -\Omega \cot (\Omega s) )^2
\right\} \Delta(\omega,\infty).
\label{aklinfty}
\end{eqnarray}
Using these results we can compute the asymptotic properties of
${\bf Q}$ needed in the calculations presented in the text.

\section{Effective spectral density}\label{app:LRT}

In this Appendix we derive the relation between the effective
spectral density for the $CB$ system and the dynamical
susceptibility of subsystem $C$ using an argument that by-passes
the actual diagonalization procedure. A similar argument is given
in Refs.~\cite{Leggett84,GOA85}. From the Hamiltonian given in
eq.~(\ref{H}), we can write the equation of motion for $C$ with a
harmonic potential in the form,
\begin{eqnarray}
x(\omega) = \chi_{C}(\omega) F_C(\omega)
\label{xchif}
\end{eqnarray}
in the complex Fourier representation with a susceptibility
function $\chi_{C}(\omega)$. The force $F_C(\omega)=- \partial
V_{AC}(\omega)/\partial x= - \lambda \sigma(\omega)$ is the
external force from subsystem $A$ acting on $C$. On the other
hand, the equation of motion for subsystem $A$ gives
\begin{eqnarray}
- m_A \omega^2 \sigma(\omega) - \lambda x(\omega) = F_A(\omega)
\label{EOMforsigma}
\end{eqnarray}
with $F_A(\omega)=- \partial V_{A}(\sigma) / \partial \sigma$.
Here we assumed $H_{A}$ has the form $p_A^2/2m_A + V_A(\sigma)$
for simplicity. Our argument does not depend on the form of
$H_{A}$ as clear from the context. Combining with
eq.~(\ref{xchif}), we have
\begin{eqnarray}
\left[ - m_A \omega^2 - \lambda^2 \chi_{C}(\omega)
\right]\sigma(\omega)  = F_A(\omega).
\label{chisigmaf}
\end{eqnarray}

Now, suppose we have already diagonalized last three terms of
eq.~(\ref{H}) and replaced them with $N+1$ - harmonic oscillators,
\begin{eqnarray}
H_C + V_{CB} + H_B \rightarrow H'_{B} =
\sum_{n=0}^{N}( \frac{p_{n}^{'2}}{2 m'_{n}}
    + \frac{m'_{n} \omega_{n}^{'2} q_{n}^{'2}}{2}).
     \label{HB'}
\end{eqnarray}
We write the interaction term $V_{AC}$ in terms of new coordinates
$q'_{n}$
\begin{eqnarray}
V_{AC} \rightarrow V'_{AB} = \sigma \sum_{n=0}^{N} c'_{n} q'_{n}.
\label{ACnew}
\end{eqnarray}
Assuming that the system described by the Hamiltonian in
eq.~(\ref{H}) reaches thermal equilibrium in the asymptotic limit,
we can replace the effect of the combined system $CB$ by the
equivalent thermal bath $H'_{B}$ at the same temperature. Although
this diagonalization procedure is straightforward, for our
purpose, we only need the form of the equation of motion for
$\sigma$ expressed by the dissipation coefficient
$\gamma_{CB}(\omega)$ for the effective bath eq.~(\ref{HB'}),
\begin{eqnarray}
\left[ - m_A \omega^2 - i m_A \omega  \gamma_{CB}(\omega)
\right]\sigma(\omega)  = F_A(\omega).
\label{FAomega}
\end{eqnarray}
Comparing the above with eq.~(\ref{chisigmaf}) we have
\begin{eqnarray}
 \lambda^2 \chi_{C}(\omega)
= i m_A \omega \gamma_{CB}(\omega).
\label{chifromgammaCB}
\end{eqnarray}
Recall that the spectral density is related to the dissipation
coefficient by
\begin{eqnarray}
  J_{CB}(\omega)  =  m_A \omega \gamma'_{CB}(\omega),
\label{JCBtogamma}
\end{eqnarray}
for real $\omega$. From this the desired relation follows:
\begin{eqnarray}
  J_{CB}(\omega)  =  \lambda^2 \chi''_{C}(\omega).
\label{JCBtoChiC}
\end{eqnarray}

\end{multicols}
\end{document}